\documentclass[a4paper,11pt]{article}
\pdfoutput=1 % if your are submitting a pdflatex (i.e. if you have
\usepackage{jcappub} % for details on the use of the package, please
\usepackage[T1]{fontenc} % if needed

\usepackage[utf8]{inputenc}
\usepackage{amsmath, amsthm, mathtools}
\usepackage[numbers,sort&compress]{natbib}

\title{\boldmath On the cosmic-ray energy scale of the LOFAR radio telescope}

\author[1]{K.~Mulrey}
\author[1,2]{S.~Buitink}
\author[1,2]{A.~Corstanje}
\author[2,3,4]{H.~Falcke}
\author[5]{B.~M.~Hare}
\author[1,2,3]{J.~R.~H\"{o}randel}
\author[1,6]{T. Huege}
\author[1]{G.~K.~Krampah}
\author[1]{P.~Mitra}
\author[7,8]{A.~Nelles}
\author[1]{H.~Pandya}
\author[1]{J.~P.~Rachen}
\author[9]{O.~Scholten}
\author[2,4]{S.~ter~Veen}
\author[10]{S.~Thoudam}
\author[11]{T.~N.~G.~Trinh}
\author[12]{T.~Winchen}

\affiliation[1]{\textit{Astrophysical Institute, Vrije Universiteit Brussel, Pleinlaan 2, 1050 Brussels, Belgium}}

\affiliation[2]{\textit{Department of Astrophysics/IMAPP, Radboud University, P.O. Box 9010, 6500 GL Nijmegen, The Netherlands}}
\affiliation[3]{\textit{Nikhef, Science Park 105, 1098 XG Amsterdam, The Netherlands}}
\affiliation[4]{\textit{Netherlands Institute of Radio Astronomy (ASTRON), Postbus 2, 7990 AA Dwingeloo, The Netherlands}}

\affiliation[5]{\textit{University of Groningen, Kapteyn Astronomical Institute, Groningen, 9747 AD, Netherlands}}

\affiliation[6]{\textit{Institut f\"{u}r Kernphysik, Karlsruhe Institute of Technology (KIT), P.O. Box 3640, 76021, Karlsruhe, Germany}}
\affiliation[7]{\textit{DESY, Platanenallee 6, 15738 Zeuthen, Germany}}
\affiliation[8]{\textit{ECAP, Friedrich-Alexander-University Erlangen-N\"{u}rnberg, 91058 Erlangen, Germany}}

\affiliation[9]{\textit{Interuniversity Institute for High-Energy, Vrije Universiteit Brussel, Pleinlaan 2, 1050 Brussels, Belgium}}

\affiliation[10]{\textit{Department of Physics, Khalifa University, PO Box 127788, Abu Dhabi, United Arab Emirates}}
\affiliation[11]{\textit{Department of Physics, School of Education, Can Tho University Campus II, 3/2 Street, Ninh Kieu District, Can Tho City, Vietnam}}
\affiliation[12]{\textit{Max-Planck-Institut f\"{u}r Radioastronomie, Auf dem Hügel 69, 53121 Bonn}}

\emailAdd{kmulrey@vub.be}
\abstract{Cosmic rays are routinely measured at LOFAR, both with a dense array of antennas and with the LOFAR Radboud air shower Array (LORA) which is an array of plastic scintillators.  In this paper, we present two results relating to the cosmic-ray energy scale of LOFAR.  First, we present  the reconstruction of cosmic-ray energy using radio and particle techniques along with a discussion of the event-by-event and absolute scale uncertainties.  The resulting energies reconstructed with each method are shown to be in good agreement, and because the radio-based reconstructed energy   has smaller uncertainty on an event-to-event basis, LOFAR analyses will use that technique in the future.  Second, we present the radiation energy of air showers measured at LOFAR and demonstrate how radiation energy can be used to compare the energy scales of different experiments.  The radiation energy scales quadratically with the electromagnetic energy in an air shower, which can in turn be related to the energy of the primary particle.  Once the local magnetic field is accounted for, the radiation energy allows for a direct comparison between the LORA particle-based energy scale and that of the Pierre Auger Observatory.  They are shown to agree to within (6$\pm$20)\% for a radiation energy of 1 MeV, where the uncertainty  on the comparison is dominated by the antenna calibrations of each experiment.  This study motivates the development of a portable radio array which will be used to cross-calibrate the energy scales of different experiments using radiation energy and the same antennas, thereby significantly reducing the uncertainty on the comparison.
}

\usepackage{soul,xcolor}
\begin{document}
\setstcolor{blue}
\maketitle
\flushbottom

\section{Introduction}

One of the main challenges in the field of cosmic-ray astrophysics is accurately determining the energy of detected cosmic rays.  Experiments use different detection, calibration, and reconstruction techniques, resulting in different energy scales.  Currently, it is necessary to shift the energy scales of different experiments in order to align them to produce an overall spectral fit~\cite{Hoerandel:2002yg,Dembinski:2017zsh}.  Understanding an experiment's energy scale is critical for comparing the spectrum and composition measurements of different experiments and for building global models of cosmic-ray sources, acceleration and propagation~\cite{Bluemer:2009zf,Deligny:2020gzq}.

An important factor that influences an experiment's energy scale and the associated uncertainties is the way in which cosmic rays are detected.  Above roughly 100~TeV, cosmic rays are detected indirectly, via the air shower that is generated when the primary cosmic ray interacts in the atmosphere.  Air shower particles that reach the Earth's surface can be sampled with detectors on the ground and used to estimate the energy in the air shower~\cite{Schieler:2003tpo,IceCube:2012nn}.  However, this method only captures a snapshot of the shower development and relies heavily on hadronic interaction models for the interpretation of data, which introduce large systematic uncertainties at high energies~\cite{Engel:2011zzb}.  An alternative approach is to measure the energy deposition from the electromagnetic part of the air shower during its longitudinal development, which eliminates most of the dependency on hadronic interaction models.  This can be done, for example, by measuring the fluorescence light emitted by air molecules that have been excited by the shower particles.  The fluorescence detection method provides a calorimetric energy measurement, but can only be done during dark nights and requires good knowledge of atmospheric conditions~\cite{2015172,Abraham:2009bc,KAWAI2008221,Abreu:2012oza}.

Air showers are also detected using the broadband radio emission that is generated as the shower develops~\cite{Falcke:2005tc,Bezyazeekov:2015rpa,schellart2013,Aab:2015vta,Ardouin:2005qe}.  The dominant contribution to the radio emission comes from the geomagnetically induced, time-varying transverse current that develops as the shower propagates~\cite{kahn:1966,Werner:2007kh}.  The strength of this emission scales with the absolute value of the geomagnetic field and the sine of the angle between the shower velocity and the geomagnetic field.  A secondary contribution comes from emission that results from the development of a charge excess in the shower front~\cite{askaryan:1962}.  Radio emission is produced primarily by the electromagnetic components of the shower, and is calculated from first principles using classical electrodynamics~\cite{huege2016}.  The measured radio signal is integrated over the whole air shower, and so measurements can be used to perform complete calorimetric energy reconstructions without having to consider absorption or scattering~\cite{Aab:2016eeq,Glaser:2016qso}.  Furthermore, the total energy radiated by the air shower in the form of radio emission, or radiation energy, of the shower can be determined by integrating the radio energy fluence (energy per unit area) footprint on the ground. Once the radiation energy is corrected for the strength of the local magnetic field and second order effects such as the effect of atmospheric density on the shower development and the relative charge excess contribution, it becomes a universal quantity.  If the radiation energy from an air shower is found in conjunction with the shower energy, determined using an independent method, it can be used to compare the energy of cosmic rays detected at different locations.  The radio measurement technique also allows for the precise reconstruction of the atmospheric depth of shower maximum, $X_{\mathrm{max}}$~\cite{buitinkNature2016,Bezyazeekov:2015ica}.  The interpretation of this parameter relies on having an accurate energy determination.

The LOw Frequency ARray (LOFAR) is a distributed radio telescope with a dense antenna array in the Netherlands~\cite{LOFAR}.  An in-situ particle detector array, the LOFAR Radboud Air Shower Array (LORA), is used to trigger antenna readout~\cite{thoudam2014}.  Each event is simultaneously sampled by the LOFAR antennas, which measure emission in the \mbox{$30-80$}~MHz band, and LORA scintillators which detect particles reaching ground level.  Features of the primary cosmic ray are reconstructed with high precision~\cite{buitink2014,schellart2013}.  In this paper we present two results related to the cosmic-ray energy scale of LOFAR.

First, we present energy reconstruction techniques using  the radio emission measured with the LOFAR antennas and particle data measured with the LORA scintillators.  The techniques are based on CoREAS and CORSIKA simulations, respectively, and provide unique energy reconstructions.  We establish that the energies reconstructed with both methods are consistent.  Until recently, within LOFAR, shower properties like $X_{\mathrm{max}}$ and energy were reconstructed using a hybrid method that included both particle and radio data, with the particle data determining the absolute energy.  In general, the event-by-event uncertainties on radio-based energy reconstructions are smaller than those of particle-based energy reconstructions, and so with this work we move to using radio measurements to set the LOFAR energy scale~\cite{Corstanje:2019}.

Second, we determine the radiation energy of each LOFAR event using a relation derived from CoREAS simulations~\cite{Huege:2013vt,Glaser:2016qso} and demonstrate how radiation energy can be used to compare the energy scales of different experiments.  Radiation energy is measured by the Auger Engineering Radio Array (AERA)~\cite{Aab:2016eeq,Schulz:2015mah} in conjunction with traditional cosmic-ray measurements made at the Pierre Auger Observatory.  This allows us to compare the LORA and Auger energy scales via the radiation energy measured at each location. %At a radiation energy of 1 MeV, the LORA and Auger energy scales are shown to agree to within (6$\pm$20)\%, where the uncertainty is dominated by the antenna calibrations of each experiment.  This study motivates the development of a portable radio array which will be used to cross-calibrate the energy scales of different experiments using radiation energy and the same antennas and detection system, thereby significantly reducing the uncertainty on the comparison.

This paper is organized as follows.  In Section~\ref{sec:lofar} the LOFAR telescope and cosmic-ray data processing techniques are introduced.  In Section~\ref{sec:energy_recon} both the radio-based and particle-based energy reconstruction methods and experimental uncertainties are described. In Section~\ref{sec:compareLOFAR}  a comparison is made of the energy reconstructions resulting from each method.  In Section~\ref{sec:radenergy} we find the radiation energy for events measured at LOFAR.  Finally, in Section~\ref{sec:compare2}, the LORA and Auger energy scales are compared using the radiation energy, and a new technique is introduced that will be used to compare energy scales between experiments in the future.

\section{The cosmic-ray energy scale of LOFAR}\label{sec:LOFARall}

This section contains a discussion of the cosmic-ray measurements made at the LOFAR telescope.  Radio and particle-based energy reconstruction techniques are described, and a comparison is made between the two.

\subsection{Cosmic-ray measurements with LOFAR}\label{sec:lofar}
LOFAR is a radio telescope with antenna stations distributed across northern Europe.  A dense core of 24 stations is located in the North of the Netherlands~\cite{LOFAR}.  At the center is the ``Superterp,'' consisting of six stations located within a 160~m radius.   Each station consists of 96 dual-polarized Low Band Antennas (LBAs) that operate in the $30-80$~MHz band as well as High Band Antennas which are not used in this analysis. The 96 LBAs are organized in inner and outer sets of 48 each, and at any given time one of the two sets is operational.  Each antenna is digitized at 200 mega-samples per second and the data are stored in a 5~s ring buffer.  Also on the Superterp is LORA, a particle detector array consisting of 20 plastic scintillators.  LORA detects showers above $10^{16}$~eV and acts as a trigger for radio readout.   When a cosmic ray is detected, the ring buffers are frozen and 2.1~ms of data are saved.  Data from the LBAs are processed offline~\cite{schellart2013} where radio frequency interference (RFI) is removed and the data are calibrated~\cite{Mulrey:2019vtz}.  The arrival direction of the event is found using the timing of arrival of the radio signal.  The voltage at each antenna position is integrated over a 55~ns time window centered at the pulse peak, resulting in what will be referred to as ``measured energy,'' $\varepsilon$.  \\
\indent The interpretation of both radio and particle data depends on the detector calibration, or in other words, how the measured signals in analog-digital conversion (ADC) units are translated into a quantity with physical meaning.  This requires comparing measured data to a known source.  The LOFAR system response, including the antennas and signal chain, is calibrated using Galactic emission as a source, as it is the primary contributor to the background in the antenna signals.  However, there is a secondary contribution to the signal from electronic noise introduced in the signal chain. In order to estimate this contribution, modeled Galactic emission is propagated through the antenna and signal chain component by component, including frequency dependent gains and losses in the system and electronic noise where it enters the system.  The electronic noise is not known a priori, and is determined using a fitting procedure that makes use of the variation of the Galactic emission measured by the LOFAR antennas as a function of local sidereal time.  The calibration can then be found by comparing measured background signals to the predicted Galactic and electronic noise signals.  The resulting calibration has an uncertainty of 13\%, which is dominated by the uncertainty on the underlying models used to predict the Galactic emission.  Details of the LOFAR system response calibration are given in~\cite{Mulrey:2019vtz}.\\
\indent The energy deposited by the shower particles in the scintillators is determined by calibrating the scintillators using single muons~\cite{thoudam2014}.  The scintillators are operated in a mode with a low trigger setting, so that singly charged particles trigger the detector readout.  In order to calculate the total signal produced by a single muon, which corresponds to the total energy deposit, the ADC time trace is integrated in a time window of $(-75, +875)$~ns around the peak of the signal. By collecting many muons, we build a distribution of energy deposits from which the most probable value can be determined.  This procedure can be done in the field, so that the scintillators are calibrated under realistic operating conditions.  Energy deposits from single muons are also simulated with GEANT4~\cite{geant4}, using an all-sky cos$^2$($\theta$) zenith angle distribution of arrival directions and a realistic description of the detector.  This provides a distribution of simulated energy deposits from which we determine the most probable value.  By comparing the peak of the measured muon distribution to the simulated muon distribution, we arrive at a calibration factor for the LORA scintillators.  More details on the scintillator calibration are provided in Appendix~\ref{sec:uncertainties}.\\
\indent A preliminary radio-based energy estimate is made using a two dimensional LDF that makes use of the asymmetry in the radio footprint due to the interference of geomagnetic and charge excess components~\cite{nelles2015}.  The radio fluence at the position of each antenna is determined, and the LDF is fit using a minimization procedure.  This results in initial estimates for energy, $X_{\mathrm{max}}$, and core position.  These values are the starting points for the final shower reconstruction which is described in the following section.

\subsection{Energy reconstruction}\label{sec:energy_recon}

The LOFAR energy reconstruction is based on CORSIKA~\cite{corsika} and CoREAS~\cite{Huege:2013vt} simulations.  CORSIKA~7.7100 is used with hadronic interaction models FLUKA~\cite{fluka} and \linebreak \mbox{QGSJETII-04}~\cite{qgsjet}.  A thinning method, where only a sub-sample of particles are tracked to make large simulations more feasible, is applied at the $10^{-6}$ level with optimized weight limitation~\cite{Kobal:2001jx}.  A GEANT4~\cite{geant4} simulation of the scintillator panels is used to derive the energy deposit from the CORSIKA-simulated particles at ground level as a function of radius from shower core, which is then used to determine the simulated energy deposit at the position of the LORA scintillators.  The `gdastool' CORSIKA plug-in is used to simulate each event with a realistic atmosphere~\cite{Mitra:2020}. The CORSIKA ``STEPFC'' parameter, which controls the electron multiple scattering length used in the EGS4 package (which handles electromagnetic interactions), is set to its default value of 1.  This value was shown to produce radiation energy 11\% lower than an optimized value of STEPFC=0.05~\cite{Gottowik:2017wio}.  In order to correct for this, we increase the simulated radiation energy by 11\%.  Details of the simulation procedure can be found in~\cite{buitink2014}.

A set of simulations is generated for each shower using the initial energy estimate and the direction determined from radio signal arrival times as input parameters.  The set contains both iron nuclei and protons as primaries and $X_{\mathrm{max}}$ values spanning the natural range. Antenna positions are simulated in a star-shaped pattern with 8 arms of 20 antennas spaced 25~m apart.  The pattern is generated in the shower plane and then projected onto the ground plane, so that two arms align with the $\mathbf{v}\times\mathbf{B}$ axis and two with the $\mathbf{v}\times\mathbf{v}\times\mathbf{B}$ axis, where $\mathbf{v}$ is the velocity axis of the shower front and $\mathbf{B}$ is the Earth's magnetic field. 

The simulated electric field traces are converted into voltages by applying the LBA antenna model, and bandpass filtered to $30-80$~MHz.  As is done with measured data, the simulated voltage traces are integrated over a 55~ns time window centered at the pulse peak, resulting in a simulated ``measured energy,'' $\varepsilon_{\mathrm{sim}}$.  Radio and particle-based energy reconstructions are then done using $\chi^2$ fitting procedures, which are described in the following subsections.

\subsubsection{Radio energy reconstruction}\label{sec:energy_radio}

\begin{figure}
\centering
\includegraphics[scale=0.43]{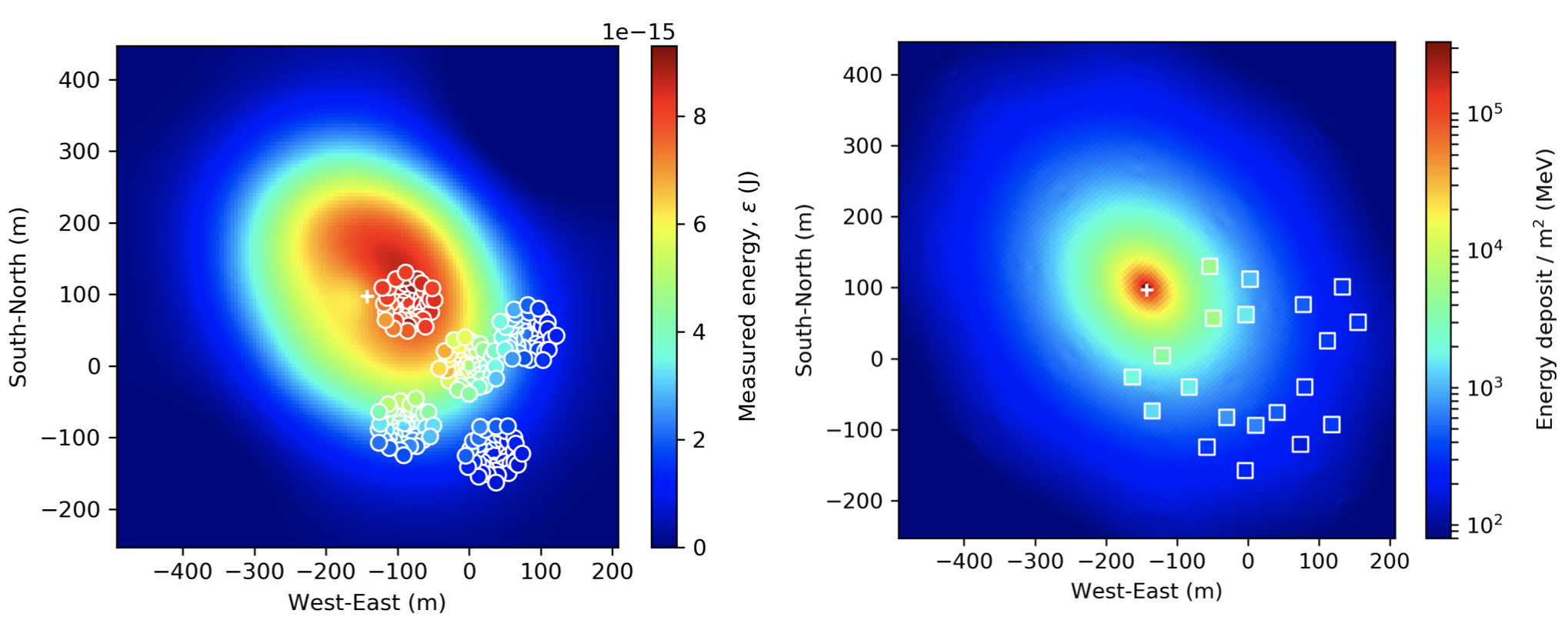}
\caption{Left: Example radio footprint for a LOFAR event, shown in the ground plane, with energy $7.2 \times 10^{17}$~eV, zenith angle $37^{\circ}$, azimuth angle $41^{\circ}$ south of east, and $X_{\mathrm{max}}=670$~g/cm$^2$.  The color scale represents measured energy, $\varepsilon$, at  each position.  The background map results from simulations, and the colors in the white circles represent data measured with the LOFAR antennas.  Right: Map of the particle energy deposit in LORA scintillators for the same event.  The color scale represents the energy deposit per square meter in MeV.  The background color comes from simulations, and the filled squares represent energy deposits measured by LORA scintillators.}
\label{fig:reco}
\end{figure}

The radio-based energy reconstruction is done by comparing $\varepsilon_{\mathrm{sim}}$ at each LOFAR antenna position to the detected $\varepsilon$.  For this, a two-dimensional radiation map is generated by interpolating $\varepsilon_{\mathrm{sim}}$ in the antennas simulated in the star-shaped pattern in the shower plane.  This way, $\varepsilon_{\mathrm{sim}}$ at any given antenna position $(x_{\mathrm{ant}},y_{\mathrm{ant}})$ can be obtained.  The map is then fit to LOFAR data using a minimization procedure with free parameters for the core position of the shower and a scale factor for the energy, as
\begin{equation}\label{eq:chi2}
    \chi^2_{\mathrm{radio}}=\sum_{\mathrm{antennas}}\bigg( \frac{\varepsilon-f_r^2\varepsilon_{\mathrm{sim}}(x_{\mathrm{ant}}-x_0,y_{\mathrm{ant}}-y_0)}{\sigma_{\mathrm{ant}}}  \bigg)^2
\end{equation}
where $\sigma_{\mathrm{ant}}$  refers to the one sigma level of the time-integrated voltage of measured traces 
outside the signal window, ($x_0,y_0$) is the shower core position, and $f_r$ is the energy scaling factor, allowing for deviations from the simulated cosmic-ray energy.  A $\chi^2$ fit is done for each simulation in the set.  This is similar to the fit procedure that has been used for LOFAR $X_{\mathrm{max}}$ analyses in the past~\cite{buitink2014,buitinkNature2016}, the difference being that here there is no particle data information included in the fit.  In the case that the scale factor is greater than $2$ or less than $0.5$, a new set of simulations is run with a new starting energy estimate.  The values of $f_r$, $x_0$, and $y_0$ for the simulation with minimum $\chi^2$ are taken as the reconstructed values for this event, and the radio-based energy is found as
\begin{equation}\label{eq:radio_energy}
    E_{\mathrm{radio}} = f_r \times E_{\mathrm{sim}}.
\end{equation}
An example of the best fit $\varepsilon_{\mathrm{sim}}$ map to measured data is shown in the left panel of Fig.~\ref{fig:reco}.  The background color represents the $\varepsilon_{\mathrm{sim}}$ that would be received by an antenna at each position, and the colors in the white circles represent $\varepsilon$ at specific antenna positions.  The white cross represents the shower core.

We consider systematic uncertainties, which affect the absolute scale of the reconstructed cosmic-ray energy, and event-by-event  uncertainties, both of which are outlined in Table \ref{tab:radio_uncertainties}.  An overview of the uncertainties is given here, and an extended discussion can be found in Appendix~\ref{sec:uncertainties}.  Event-by-event uncertainties on the radio-based energy include the following.  The simulated antenna model predicts the antenna response as a function of direction, which is different for each event.  By offsetting the direction of the incoming cosmic ray by $\pm5^{\circ}$ in the zenith direction and propagating the effects through the energy reconstruction process, we conservatively estimate that the uncertainties in the antenna response have at most a 5\% effect on the reconstructed energy.  Change in antenna gain as a function of temperature is found to be negligible.   We use realistic atmospheres on an event-by-event basis for our simulations, and so any effects arising from incorrect atmospheric conditions are also negligible.  There is uncertainty which comes from the event reconstruction procedure, and is estimated using a Monte Carlo vs. Monte Carlo method.  This uncertainty includes the effects of the noise levels and geometry on the reconstruction of a particular event.  Values range from 4\% to 18\%, with a typical value being 9\%.  Additionally, there is an uncertainty due to the fact that the type of the primary is unknown.  Given the same shower geometry and $X_{\mathrm{max}}$, an air shower initiated by a proton primary yields a reconstructed energy consistently 10\% lower than an air shower initiated by an iron primary.  In the fitting procedure the best-fit simulation that is used to reconstruct the energy is associated either with a proton or an iron primary.  Therefore, we add an asymmetric event-by-event uncertainty to each event to account for the unknown primary.  The typical total event-by-event uncertainty is then~14\%.

Systematic uncertainties which affect the absolute radio-based energy scale include the calibration of the antenna and signal chain, the effects of hadronic interaction models, and the choice of radio simulation package, namely, CoREAS~\cite{Huege:2013vt} or ZHAireS~\cite{AlvarezMuniz:2011bs}.  The system calibration is the dominant factor, at 13\%~\cite{Mulrey:2019vtz}.  The choice of simulation package has a 2.6\% effect on energy, and the effect of the choice of hadronic interaction model is 3\%~\cite{Gottowik:2017wio,James:2010vm}.  Although the radio emission is primarily generated by the electromagnetic component of the air shower, the choice of hadronic interaction model influences how much of the primary particle energy goes into that component.

\begin{table}
\caption{Summary of the uncertainties in the radio-based energy scale.  The $\bigoplus$ symbol indicates quadratic addition.  Details of how these uncertainties are derived can be found in Appendix~\ref{sec:uncertainties}.}
\label{tab:radio_uncertainties}
\begin{center}
 \begin{tabular}{l|c}
 
 \textbf{Uncertainty} & \textbf{Value} \\ [0.5ex] 
 \hline\hline
 \textbf{Event-by-event} & \\
 \hline
 \hspace{1cm} angular dependence of antenna model & 5\%\\
 \hline
  \hspace{1cm} temperature dependence & negligible\\
 \hline
  \hspace{1cm} reconstruction uncertainty & typically 9\%\\
 \hline
   \hspace{1cm} composition uncertainty & 10 \%\\
 \hline
\textbf{Total event-by-event} & 11\% $\bigoplus$ reconstruction uncertainty\\
\noalign{\vskip 0.5cm}    
\hline\hline

\textbf{Absolute scale} & \\
 \hspace{1cm} antenna calibration and system response & 13\% \\
\hline
 \hspace{1cm} hadronic interaction models & 3\%\\
\hline
 \hspace{1cm} radio simulation method & 2.6\% \\
\hline

\textbf{Total absolute scale} & 13.6\%\\
 
\end{tabular}
\end{center}
\end{table}

\begin{figure}
\centering
\includegraphics[scale=0.50]{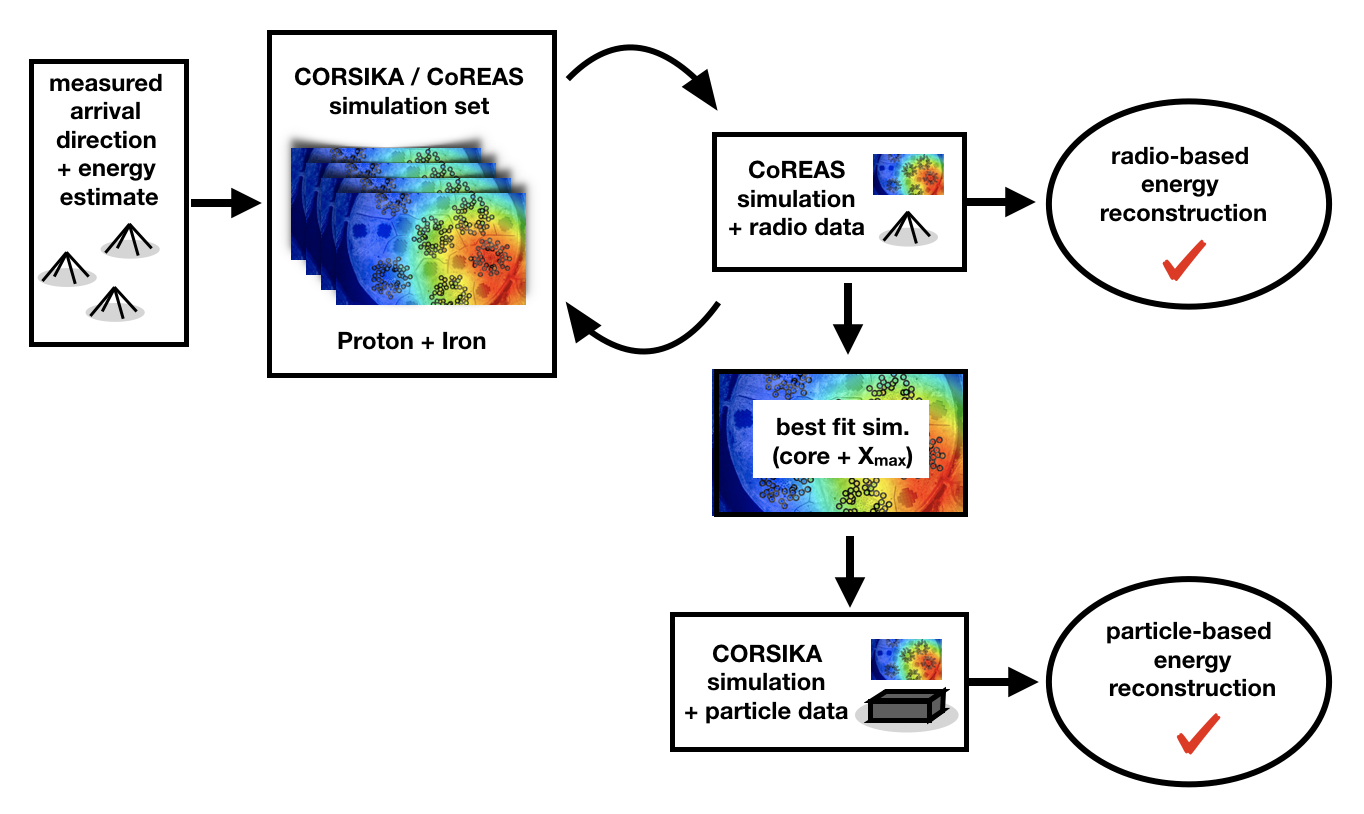}
\caption{Diagram of the radio and particle energy reconstruction procedures.  The radio-based reconstruction is done first, resulting in a best-fit simulation. This is then used in the particle-based reconstruction.}
\label{fig:energy_flow}
\end{figure} 

\subsubsection{Particle energy reconstruction}\label{sec:energy_particle}

The particle-based energy reconstruction uses the best fit CORSIKA simulation as determined by the radio $\chi^2$ fit.  The radio-based measurements are more sensitive to $X_{\mathrm{max}}$ and shower development, and so are more reliable for determining the best simulation.  Furthermore, the core position from the radio measurement is used.  When the core position falls outside the footprint of the particle detectors, it is difficult to constrain.  In this sense, the two reconstructions are not entirely independent, but once the shower geometry is fixed the scale set for the particle-based energy is determined entirely by the scintillator data. A diagram of the radio and particle energy reconstruction procedures is shown in Figure~\ref{fig:energy_flow}.

Because we use results from the radio reconstruction to fix the shower geometry, many of the typical uncertainties associated with particle-based reconstructions are avoided.  We take this approach, and do not attempt a completely independent particle-based reconstruction for two reasons.  First, in past LOFAR analyses, a hybrid method was used to reconstruct energy which combined radio and particle information.  The radio information was primarily used to determine the event geometry, while the energy was set using the particle information.  The particle-based method described in this section is consistent with past analyses, and therefore allows us to compare the energy set in past analyses with the new radio-based technique.  Secondly, because of the small footprint of the scintillator array, independent particle-based reconstructions would suffer from poor core reconstruction, which would severely reduce the number of usable events we have, and furthermore, the resulting uncertainties would be prohibitively large.  It should be noted that LORA detects a large number of lower energy air showers that fall within the footprint of the scintillator array but do not have a strong radio signal.  In this work we are interested in comparing the radio and particle reconstruction techniques, so these events aren't considered, however they were the source of an independent spectral analysis~\cite{thoudam2016}.

With the best simulation known, a particle $\chi^2$ is fit, as
\begin{equation}
    \chi^2_{\mathrm{particle}}=\sum_{\substack{\mathrm{particle}\\ \mathrm{detectors}}}\left( \frac{d_{\mathrm{det}}-f_pd_{\mathrm{sim}}}{\sigma_{\mathrm{det}}} \right)^2
\end{equation}
where $d_{\mathrm{det}}$ is the
deposited energy measured by a LORA detector with one standard deviation of background noise level $\sigma_{\mathrm{det}}$, and $d_{\mathrm{sim}}$ is the GEANT4 simulated deposit.  The particle scale factor $f_p$ is needed to bring the simulated energy into agreement with the measurements.  An example of the particle footprint of the best fit simulation is shown in the right panel of Fig.~\ref{fig:reco}.  The background color represents the simulated energy per square meter that would be deposited in a LORA scintillator at each location.  On average, one muon deposits~6.5~MeV in a scintillator panel.  The squares represent the locations of the scintillators with the color being the measured energy deposit. Again, the white cross represents the shower core.  The particle-based energy is then found as
\begin{equation}
\begin{split}
    E_{\mathrm{particle}} &= f_p \times E_{\mathrm{sim}}\\&= \frac{f_p}{f_r} \times E_{\mathrm{radio}}.
\end{split}
\end{equation}

The uncertainties on the particle-based energy reconstruction are summarized in Table~\ref{tab:particle_uncertainties}.  Event-by-event uncertainties include variation in the detector response, as well as the reconstruction uncertainty for each event.  Scintillators have on average a 10\% fluctuation in response, but since the fluctuations are not correlated between scintillators, this propagates into a 2.5\% uncertainty on the reconstructed energy.  The reconstruction uncertainty on the particle-based energy, like the radio-based energy, is determined using a Monte Carlo vs. Monte Carlo method.  The reconstruction uncertainty on the particle-based energy is larger than that of the radio-based energy, and extends from 10\% up to 50\%.  The unknown composition of the measured events also contributes a significant uncertainty.  Although constraining the shower geometry using information from the radio fit reduces this uncertainty, we find that for events with the same $X_{\mathrm{max}}$ and geometry, the reconstructed energies for showers initiated by iron primaries are lower than for proton-initiated showers.  This effect is a function of zenith angle, with almost no difference in energy reconstructions for vertical showers, and up to 30\% difference at zenith angles around $50^{\circ}$.  We have parameterized this effect, and added an asymmetric event-by-event uncertainty accordingly, based on the primary of the best-fit simulation.

Systematic uncertainties include the uncertainty on the scintillator calibration and the effect of the choice of hadronic interaction model used in the simulations.  Using a field calibration method, scintillator calibration can be performed at any time and the uncertainty in calibration values propagates into 3\% uncertainty in energy.  In order to estimate the uncertainty introduced by the choice of hadronic interaction model, we simulated a subset of events using both Sibyll 2.3c and QGSJETII-04 and found an average 7\% difference in reconstructed energy between the two.  More details about the uncertainties on the particle energy are given in Appendix~\ref{sec:uncertainties}.

It is also known that above $10^{16}$~eV, there is a discrepancy between the number of simulated and measured muons in air showers~\cite{Cazon:2019ICRC,Arteaga:2019ICRC,Muller2019,Aab:2020muon},   with a 50\% deficit in simulated muons at $10^{17.5}$~eV reported by Auger.  In order to quantify the effect this has on our reconstructed energy, we ran simulations, artificially inflating the muonic component of the signal at ground level by 50\%.   The signal in the scintillators is dominated by the electromagnetic component of the shower, so the muon discrepancy does not have a large effect on the reconstructed energy.  For the majority of our events, increasing the simulated muonic component of the air shower by 50\% corresponded to a $\sim$5\% decrease in reconstructed energy.  At zenith angles above 40$^{\circ}$, the effects becomes larger, up to $\sim$10\%.

\begin{table}
\caption{Summary of the uncertainties in the particle-based energy scale.  The $\bigoplus$ symbol indicates quadratic addition.  Details of how these uncertainties are derived can be found in Appendix~\ref{sec:uncertainties}. Note: this technique for reconstructing the energy based on particle information also uses information from the radio reconstruction, such as shower geometry and age, reducing the overall uncertainties.}
\label{tab:particle_uncertainties}
\begin{center}
 \begin{tabular}{l|c}
 
 \textbf{Uncertainty} & \textbf{Value} \\ [0.5ex] 
 \hline\hline
 \textbf{Event-by-event} & \\
 \hline
  \hspace{1cm} scintillator response variation & 2.5\%\\
 \hline
  \hspace{1cm} reconstruction uncertainty & $10-50$\%\\
 \hline
   \hspace{1cm} composition uncertainty & $2-30$\%\\
 \hline
\textbf{Total event-by-event} & 2.5\% $\bigoplus$ reconstruction uncertainty\\
&   $\bigoplus$ composition uncertainty\\
\noalign{\vskip 0.5cm}    
\hline\hline

\textbf{Absolute scale} & \\
 \hspace{1cm} scintillator calibration & 3\% \\
\hline
 \hspace{1cm} hadronic interaction models & 7\%\\
\hline

\textbf{Total absolute scale} & 7.6\%\\
 
\end{tabular}
\end{center}

\end{table}

\subsection{Comparison of energy reconstruction techniques}\label{sec:compareLOFAR}

\begin{figure}
\centering
\includegraphics[scale=0.55]{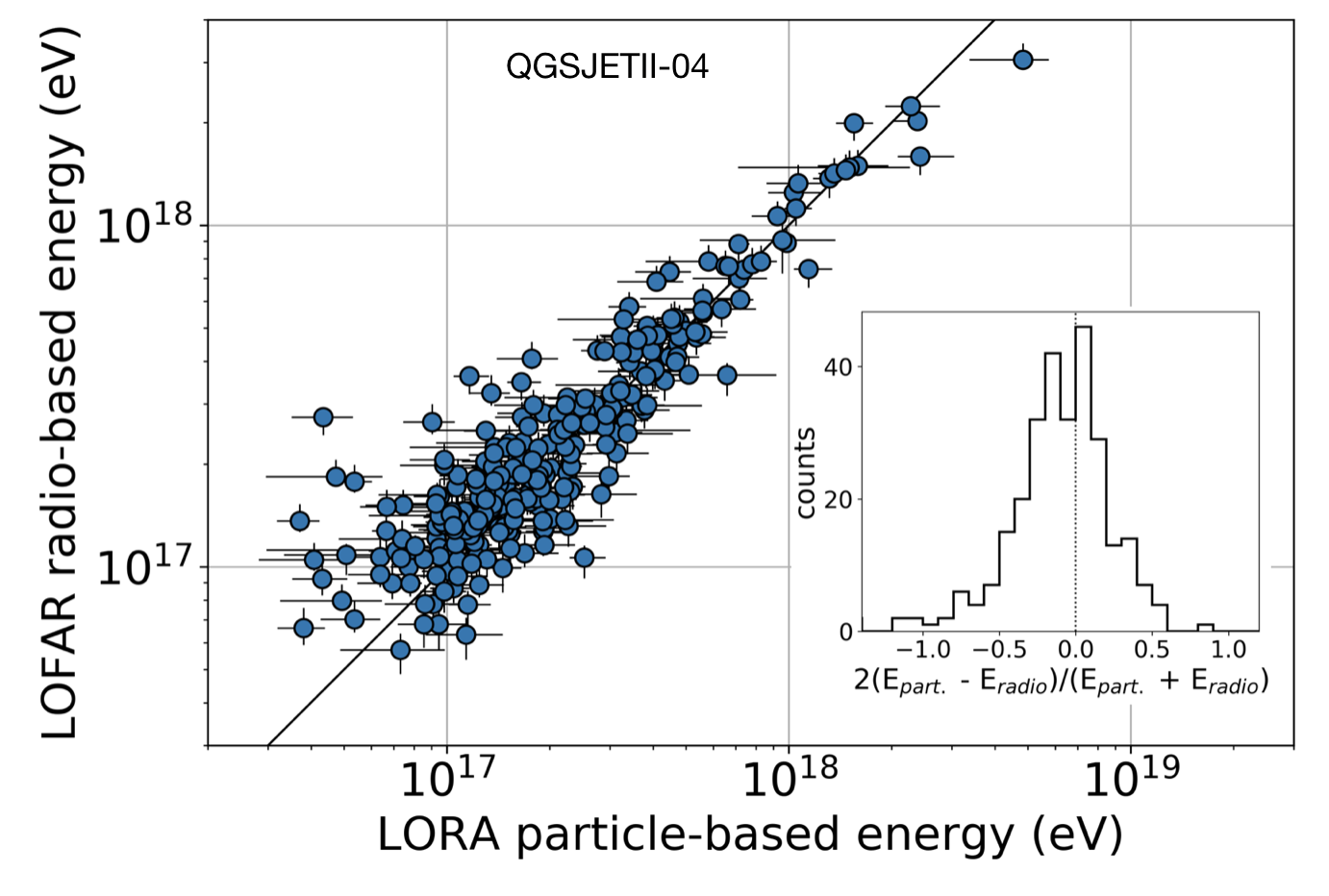}
\caption{Comparison of energy reconstructed using radio and particle-based methods.  Error bars indicate event-by-event uncertainties.  Inset: Relative difference between particle-based and radio-based reconstructed energy.  The mean of the distribution, found using a Gaussian fit, is -0.07, and the standard deviation 0.35.}
\label{fig:LOFAR_comparison}
\end{figure}

In the past, LOFAR $X_{\mathrm{max}}$ analyses used the scintillator measurements to set the energy scale. With an absolute antenna calibration available~\cite{Mulrey:2019vtz}, and because the event-by-event uncertainties on the energy reconstruction using radio data are significantly lower than for the reconstruction using particle data, LOFAR analyses will move to the radio-only approach from now on~\cite{Corstanje:2019}.  Here, we compare the radio-based and particle-based energy reconstruction of each event to demonstrate the consistency of the results.  Fig.~\ref{fig:LOFAR_comparison} shows the relation between energy reconstructed with LOFAR radio data and LORA particle data.  The error bars represent the event-by-event uncertainties of each event.  For both radio and particle-based energies, these uncertainties are dominated by the uncertainty on the reconstruction technique, discussed in Section~\ref{sec:energy_recon}.  The data consist of 283 events between 2011 and 2018, where we have chosen events where both radio and particle fits converge, the reconstructed core position has an uncertainty of less than 5~m, and the radio and particle reduced $\chi^2$ of each event are less than 5.  We also only consider events with particle-based energy reconstructed above $10^{16}$~eV.  Since the radio signal scales with energy and is thus small at low energies, the detection is biased to upward fluctuations.  The diagonal line represents a one-to-one correlation.  The inset histogram shows the relative difference between radio-based and particle-based energy reconstructions.  The mean of the distribution, found using a Gaussian fit, is -0.07, and the standard deviation is 0.35.  %In previous analyses, which used a hybrid radio-particle technique to determine the energy, the energy resolution was found to be 32\%, which is consistent with what we find here~\cite{buitinkNature2016}.%We also note that in past analyses which used a hybrid radio-particle technique to determine the energy, the event-by-event uncertainty on the energy was determined by looking at the standard deviation of the distribution of the ratio of the radio and particle scale factors, $f_r/f_p$ for all events.  This yielded an uncertainty of 32\%, which is consistent with what is presented here~\cite{buitinkNature2016}.

\noindent There is a small population of events towards the lower energies where the radio-based reconstructions are higher than the particle-based reconstructions.  Some of these events have relatively small event-by-event uncertainties compared to their distance from the line.  Normally there would be fluctuations on either side of the mean from each method.  However, since the radio signal becomes difficult to detect at low energies, it is more probable to have upwards fluctuations of radio-based energy reconstructions than downward fluctuations.  This is also evident in the asymmetry of the relative energy histogram.  As this effect is insignificant for our analysis, we do not correct for it.  We conclude that the radio and energy based reconstructions are consistent within the resolution of the measurements.

\section{Using radiation energy to compare energy scales of different experiments}\label{sec:compare}

%A powerful aspect of detecting cosmic rays using the radio method is that it allows for the determination of 
%The radiation energy scales quadratically with the energy in the electromagnetic component of the air shower. 
In this section we will demonstrate how radiation energy, $S_{RD}$, which is the total amount of radio emission emitted by an air shower, can be used to compare the energy scales of different experiments when measured in conjunction with the total shower energy, determined using an independent method.
%\subsection{Corrections to the radiation energy}\label{sec:rad_intro}
\subsection{Determination of the LOFAR radiation energy}\label{sec:radenergy}

Here, we first describe radiation energy and the corrections needed to make it a universal quantity in a general way.  Then, we will lay out how we use the results presented in~\cite{Glaser:2016qso}, which are based on CoREAS simulations, to determine the corrected radiation energy for LOFAR events.

The radiation energy contained in an air shower can be found
by integrating the energy deposit per unit area, or fluence, over the radio footprint.  Since radio waves are not attenuated in the atmosphere, the altitude of the observation site is inconsequential as long as all the radiation energy in the shower has been released before it reaches the ground.  This makes radiation energy an ideal value to compare between different locations.  However, it must still be corrected for various factors in order to make it a universal quantity which can be compared between different experiments and events.  The corrections outlined here follow those described in~\cite{Glaser:2016qso}.

The strength of the local magnetic field influences the strength of the geomagnetic emission.  For example, the magnetic field at LOFAR is 0.492~G, which is twice as strong as the 0.243~G magnetic field at Auger.  This means for the same event, the radiation energy detected at LOFAR would be larger. To correct for this, the radiation energy is scaled by the ratio of the magnetic field of a reference location and the local magnetic field.   Furthermore, when the radiation energy is calculated, the geomagnetic contribution to the radiation energy is corrected to account for the shower geometry, in particular, the angle between the magnetic field and shower axis, $\alpha$.  The charge excess fraction of the radiation energy has been parameterized so that this correction can be made to only the geomagnetic contribution.   The radiation energy also depends on the air density in which the air shower develops;  showers that develop in a less dense atmosphere have relatively larger radiation energy.  In general, one has to consider ``clipping'' effects when the radiation energy is not yet completely released at the observer height.  For LOFAR events, this is not a concern due to the fact that the observation height is close to sea level and that most shower energies are below $10^{18}$~eV.  Making these corrections yields the ``corrected radiation energy'' which can be compared between different experiments.

The work presented in~\cite{Glaser:2016qso} used CoREAS simulations to derive a relation between the corrected radiation energy in the $30-80$~MHz band, $S_{RD, \textrm{corr}}$, and the electromagnetic energy, $E_{\textrm{em}}$, contained in the air shower.  The radiation energy was found by integrating the simulated fluence in the $30-80$~MHz band over the radio footprint, and making the corrections discussed above.  The magnetic field strength was normalized to the magnetic field strength at Auger.  This resulted in the equation

\begin{equation}\label{eq:rad}
    S_{RD, \textrm{corr}}=A \times 10^7 \textrm{eV} (E_{\textrm{em}}/10^{18}  \textrm{eV})^B
\end{equation}
where $A=1.683 \pm 0.004,$ and $B=2.006 \pm 0.001$.

We use this result to directly find the corrected radiation energy of LOFAR events. In the fitting method described in Section~\ref{sec:energy_radio}, we determine the CoREAS simulation that best fits the measured radio data. Once this is known the energy contained in the electromagnetic component of the air shower, $E_{\mathrm{em}}$, can be found using the longitudinal profile information provided by the associated CORSIKA simulation, scaled by the radio scale factor $f_r$ found in Section~\ref{sec:energy_radio}. We then use Eq.~\ref{eq:rad} to determine the corrected radiation energy.  Finding $S_{RD, \textrm{corr}}$ this way is only possible because we use a simulation-based reconstruction technique and identify a best fit simulation for each event.  This gives us direct access to information about the (simulated) electromagnetic component of the air shower.  Alternatively, one can integrate the fluence over the radio footprint of the best-fit simulation and make the corrections described above to determine the corrected radiation energy.   We have confirmed that both methods yield the same result.  We also note that the fraction of $E_{\textrm{em}}$ to the total cosmic-ray energy depends on the choice of hadronic interaction model, where here we have used QGSJETII-04.

\subsection{Comparison of the energy scales of LORA and the Pierre Auger Observatory}\label{sec:compare2}

Equation~\ref{eq:rad} gives the relation between the corrected radiation energy, $S_{RD,corr}$, and the electromagnetic energy in the shower, $E_{\textrm{em}}$.  In order to compare the energy scales of different experiments using the universal quantity $S_{RD,corr}$, we need a relation between it and the total cosmic-ray energy, $E_{\mathrm{CR}}$.  
We will use a function of the same form as equation~\ref{eq:rad} to relate $S_{RD,corr}$ and $E_{\mathrm{CR}}$, written generally as $S_{RD,corr}=A'\times 10^7 \textrm{eV}(E_{\mathrm{CR}}/10^{18}eV)^{B'}$.  %This form works well because $E_{\mathrm{em}}$ scales with $E_{\mathrm{CR}}$.  
We use the notation $A',B'$ for the parameters in this case to distinguish them from $A,B$ in equation~\ref{eq:rad}, and we emphasize that they should not be directly compared because here we relate $S_{RD,corr}$ and $E_{\mathrm{CR}}$, whereas equation~\ref{eq:rad} relates $S_{RD,corr}$ and $E_{\mathrm{em}}$.  For reference, in the LOFAR energy range roughly 85\% of the total cosmic-ray energy goes into the electromagnetic components of the air shower.  The fraction of $E_{\mathrm{CR}}$ that goes into $E_{\mathrm{em}}$ differs by particle type, with protons generating showers with a higher fraction of electromagnetic particles than iron nuclei. The difference is roughly 4.5\% at 10 EeV, and decreases towards higher energies~\cite{Engel:2011zzb,Aab:2019cwj}.  Since we expect the experiments we want to compare to measure a similar composition, we don't take this difference into account.

Because the radio and particle components of each LOFAR event are measured simultaneously, we can derive the relation between $S_{RD,corr}$ determined using the LOFAR antennas and the particle-based cosmic-ray energy determined using the LORA scintillators, $E_{\mathrm{CR}^{\textrm{LORA}}}$.  In order to avoid any biases in detection that occur at lower energies, only events where both the radio and particle-based energies reconstruct to greater than $1.3\times 10^{17}~\textrm{eV}$ are used.  Using an Orthogonal Distance Regression~\cite{Boggs1989} method and fitting the equation

\begin{equation}\label{eq:LORA_rad}
    S_{RD,corr}=A'_{\mathrm{LORA}} \times 10^7 \textrm{eV} (E_{\textrm{CR}^{\mathrm{LORA}}}/10^{18}  \textrm{eV})^{B'_{\mathrm{LORA}}}
\end{equation}
results in the parameters  $A'_{\mathrm{LORA}}=1.57\pm 0.12(\textrm{stat})\pm 0.49(\textrm{sys}))$, $B'_{\mathrm{LORA}}=2.07\pm0.06(\textrm{stat})$, where the statistical uncertainty comes from the fitting procedure.  The systematic uncertainty includes both the uncertainties on the radio-based energy scale which propagate into the radiation energy, and the uncertainties on the particle-based energy scale.
%It should also be noted that parameters $A,B$ from equation~\ref{eq:rad} and $A',B'$ from~\ref{eq:LORA_rad} are not directly comparable.  Equation~\ref{eq:rad} relates corrected radiation energy to the electromagnetic energy in the air shower, and equation~\ref{eq:LORA_rad} relates corrected radiation energy to the total cosmic-ray energy.}

\begin{figure}
\centering
\includegraphics[scale=0.65]{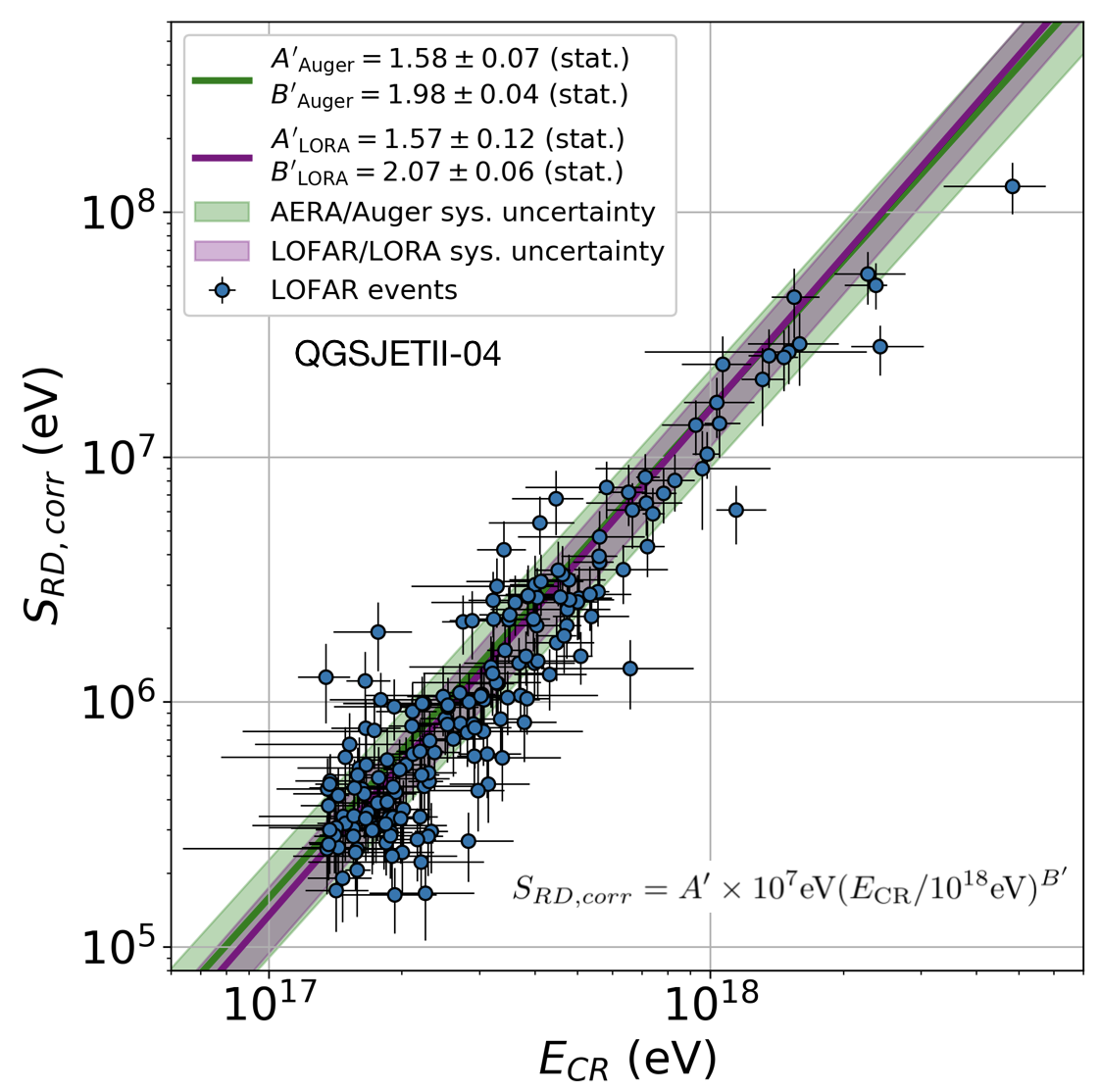}
\caption{Relation between the corrected radiation energy measured by the LOFAR antennas and the cosmic-ray energy as determined by the LORA scintillators.  The error bars represent event-by-event uncertainties.  The purple line shows the best fit line for LOFAR measurements of corrected radiation energy and LORA cosmic-ray energy, and the banded region around the best fit line represents the systematic uncertainties on the corrected radiation energy.  The green line shows the best fit line for AERA measurements of corrected radiation energy and Auger cosmic-ray energy~\cite{Aab:2016eeq}, and the shaded green region represents the systematic uncertainties on the corrected radiation energy.  QGSJETII-04 was used in the simulations on which the LOFAR and LORA energy reconstructions are based.}
\label{fig:AERAcomparison}
\end{figure}

Fig.~\ref{fig:AERAcomparison} shows the LOFAR corrected radiation energy as a function of LORA cosmic-ray energy.  Error bars indicate the event-by-event uncertainties on each point.  The resulting best-fit line of the form~\ref{eq:LORA_rad} is also shown in purple, with the shaded region indicating the absolute scale uncertainty on the corrected radiation energy.

We compare the energy scales of LORA and Auger by making use of the relation between the corrected radiation energy measured with AERA and the total cosmic-ray energy, $E_{\mathrm{CR}^{\mathrm{Auger}}}$, as determined by the Auger surface detectors~\cite{maris2011}  which are calibrated using the calorimetric energy measurements of the fluorescence detectors~\cite{Verzi2013}.  We denote the AERA corrected radiation energy as $S_{RD,corr}^{*}$, to indicate that in this case corrections are made for the relative strength of the geomagnetic emission, but second order corrections for charge excess fraction and air density in the region of shower development are not included. Derived in~\cite{Aab:2015vta} and~\cite{Aab:2016eeq}, the total corrected radiation energy in the $30-80$~MHz band can be related to total cosmic-ray energy using

\begin{equation}\label{eq:rad_auger}
    S_{RD,corr}^{*}=\frac{S_{RD}}{\mathrm{sin}^2(\alpha)}=A'_{\mathrm{Auger}} \times 10^7 \textrm{eV} (E_{\mathrm{CR}^{\mathrm{Auger}}}/10^{18} \mathrm{eV}  )^{B'_{\mathrm{Auger}}}
\end{equation}
where $A'_{\mathrm{Auger}}=1.58\pm0.07(\textrm{stat})\pm 0.67(\textrm{sys}))$, $B'_{\mathrm{Auger}}=1.98\pm0.04(\textrm{stat})$, and $\alpha$ is the angle between magnetic field and shower axis.  The systematic uncertainties include both the 16\% uncertainty (at $10^{17.5}$~eV) on the Auger energy scale~\cite{Verzi2013}, and the 14\% uncertainty on the AERA antenna calibration~\cite{Schulz:2015mah}.    Equation~\ref{eq:rad}, used to find the corrected radiation energy for LOFAR, already includes a normalization of the local magnetic field to that of Auger. Therefore the parameters $A'_{\mathrm{LORA}},B'_{\mathrm{LORA}}$ and $A'_{\mathrm{Auger}},B'_{\mathrm{Auger}}$ and in equations~\ref{eq:LORA_rad} and~\ref{eq:rad_auger} are comparable, with the caveat that second order corrections are not made for the Auger radiation energy.  Equation~\ref{eq:rad_auger} is also shown in Fig.~\ref{fig:AERAcomparison} in green, with the shaded region indicating the absolute scale uncertainties on the radiation energy. 

Using equations~\ref{eq:LORA_rad} and~\ref{eq:rad_auger}, we compare the average energies of LORA and Auger at a corrected radiation energy $S_{RD,corr}=1~\textrm{MeV}$.  This value of $S_{RD,corr}$ was chosen for the comparison because it is close to the average value of the LOFAR corrected radiation energy (determined in log-space).  $S_{RD,corr}=1~\textrm{MeV}$ corresponds to a LORA energy of $(2.64\pm0.42\mathrm{(sys)}) \times10^{17}~\textrm{eV}$ and an Auger energy of $(2.48\pm0.52\mathrm{(sys)})\times10^{17} \textrm{eV}$.  The ratio between the LORA and Auger energies $E_{\mathrm{CR}^{\mathrm{LORA}}}/E_{\mathrm{CR}^{\mathrm{Auger}}}=1.06 \pm 0.20$, meaning that  LORA energy is 6\% larger, with a $20\%$ combined uncertainty on the comparison.  The uncertainty on the comparison includes only the radio-based uncertainties on the radiation energy, namely the uncertainties on the calibration of the AERA antennas and the LOFAR radio-based energy scale.  We do not include the uncertainties associated with the Auger or LORA energy scales because we are not making a statement about the absolute energy scale of either experiment, only the comparison between the two.  The $B'_{\mathrm{LORA}}$ and $B'_{\mathrm{Auger}}$ parameters of equations~\ref{eq:LORA_rad} and~\ref{eq:rad_auger} are different, so the ratio between LORA and Auger energy changes with energy.  At the low end of the LOFAR corrected radiation energy range, $S_{RD,corr}=0.25~\textrm{MeV}$, $E_{\mathrm{CR}^{\mathrm{LORA}}}/E_{\mathrm{CR}^{\mathrm{Auger}}}=1.11 \pm 0.20$. At the high end,  $S_{RD,corr}=10~\textrm{MeV}$, $E_{\mathrm{CR}^{\mathrm{LORA}}}/E_{\mathrm{CR}^{\mathrm{Auger}}}=1.01 \pm 0.20$.  For all these points, the energy scales of Auger and LORA agree within the comparison uncertainty, although it should be noted that the comparison uncertainty is larger than the absolute uncertainty on either energy scale.  The LORA cosmic-ray energy is derived using an approach based on hadronic interaction models.  In contrast, the Auger cosmic-ray energy comes from surface detector measurements, which are calibrated using the fluorescence detectors which make model-independent, calorimetric measurements~\cite{Verzi2013}.  Nevertheless, this comparison shows that the energy scales of the two experiments are compatible, albeit with a large uncertainty on the comparison.

Radio techniques have previously been used to compare the energy scales of different experiments. The energy scales of Tunka-133~\cite{Berezhnev:2012vq} and KASCADE-Grande~\cite{KASCADE_NIM} have been compared using their radio extensions, Tunka-Rex~\cite{Schroder:2017nkf} and LOPES~\cite{Schroder:2010ZhAireSzz}.  This was done both by comparing the absolute amplitude of the radio measurements 100~m from the shower core at each location with the energies measured by Tunka-133 and KASCADE-Grande, and by using a simulation-based method~\cite{Apel:2016gws}.  This study benefited from the fact that the antennas at each location were calibrated using the same technique, which reduced the uncertainties on the comparison.  The remaining uncertainties on the comparison were due to the LOPES antenna model and the fact that the absolute amplitude of the signal at 100~m was compared, rather than radiation energy.  This value is harder to compare because it relies on knowledge of the radio footprint at a particular location, and corrections have to be made for observation level and zenith angle.  In the study presented here, we have avoided these location and shower-specific uncertainties by using the universal measurement of corrected radiation energy rather than signal amplitude.  However, the uncertainties associated with our comparison are large, because for each experiment, the uncertainties on the radiation energy, most notably the antenna calibration, must be included.

Having a method to compare the energy scales of different experiments with minimal uncertainty is necessary in order to make meaningful comparisons of their spectra and composition measurements, which are used to build global models of cosmic-ray sources, acceleration and propagation~\cite{Hoerandel:2002yg,Dembinski:2017zsh}.  We plan do this by combining the techniques used here and in~\cite{Apel:2016gws}.  A portable array of antennas will be built and deployed at various experiments, measuring radiation energy in conjunction with the host experiment's traditional air shower measurements.  The radiation energy can feasibly be reconstructed with only 5 antennas, as AERA has demonstrated~\cite{Aab:2016eeq}.  Using radiation energy to compare the energy scales eliminates uncertainties due to measurements being made at different locations, and using the same array eliminates the uncertainties associated with the antennas and calibration.  This will allow for a cross-calibration of the energy scales of different experiments with minimal uncertainty.

\section{Conclusions}\label{sec:summary}

Cosmic-ray air showers are regularly detected at LOFAR, where simultaneous measurements are made with antennas and particle detectors.  In this work we compared the reconstructed energies using radio-based and particle-based methods.  The reconstruction methods are based on CoREAS and CORSIKA simulations respectively, where simulated radio and particle footprints are fit to measured data using $\chi^2$ minimization processes.  In the past, LOFAR analyses have used a hybrid approach in which both radio and particle data were fit simultaneously.  The radio information was primarily used to determine the shower geometry, while the energy was set with the particle information.  In this work, the energies are fit separately, yielding two unique energy reconstructions.  We have shown that both methods of determining energy produce consistent results.

We discussed the uncertainties on both methods, and find a 13.6\% systematic uncertainty on radio-based energy and 7.6\% on particle-based energy.  The event-by-event uncertainties on radio-based energies are around 14\%.  For particle-based energies the event-by-event uncertainties extend to more than 50\%.  Both are dominated by the uncertainty in the event reconstruction procedure and unknown composition.  Taking into consideration the relative uncertainties, we now move from using the hybrid fitting method to using the radio-based reconstruction method to set the LOFAR energy scale.

We also used the corrected radiation energy in the $30-80$~MHz band to compare the LORA and Auger energy scales.  
Radiation energy between experiments can be compared because it represents a calorimetric energy measurement that primarily depends on the local magnetic field, which is a well known quantity and which can be accounted for.  Second order corrections can also be made with knowledge of the shower geometry and atmospheric conditions. Using the relations derived in equations~\ref{eq:LORA_rad} and~\ref{eq:rad_auger}, we determined that for a corrected radiation energy of 1~MeV the difference in LORA and Auger energies is 6\% with an uncertainty on the comparison of 20\%.

Moving forward, we will use corrected radiation energy to cross-calibrate the energy scales of different experiments.  The systematic uncertainties that are relevant for the comparison of the corrected radiation energy between experiments are dominated by the uncertainties associated with antenna and system calibrations, here 14\% for AERA antennas and 13\% for the LOFAR antenna and system response.  By using the same detection system in each location, these uncertainties can be removed from the comparison of relative energies.  We plan to build a portable array of antennas which will be used to measure radio emission from air showers in situ at different experiments.  The corrected radiation energy will then be used to directly compare the energy scales of the experiments with minimal systematic uncertainties, allowing for the establishment of a universal energy scale.

\appendix
\section{Uncertainties}\label{sec:uncertainties}

This appendix addresses how the uncertainties on the reconstructed energy of LOFAR events were derived.

\subsection{Event-by-event Uncertainties}

\vspace{4mm}
\noindent\textbf{Radio and Particle: reconstruction uncertainties}
\vspace{4mm}

\begin{figure}[ht]
\centering
\includegraphics[scale=0.45]{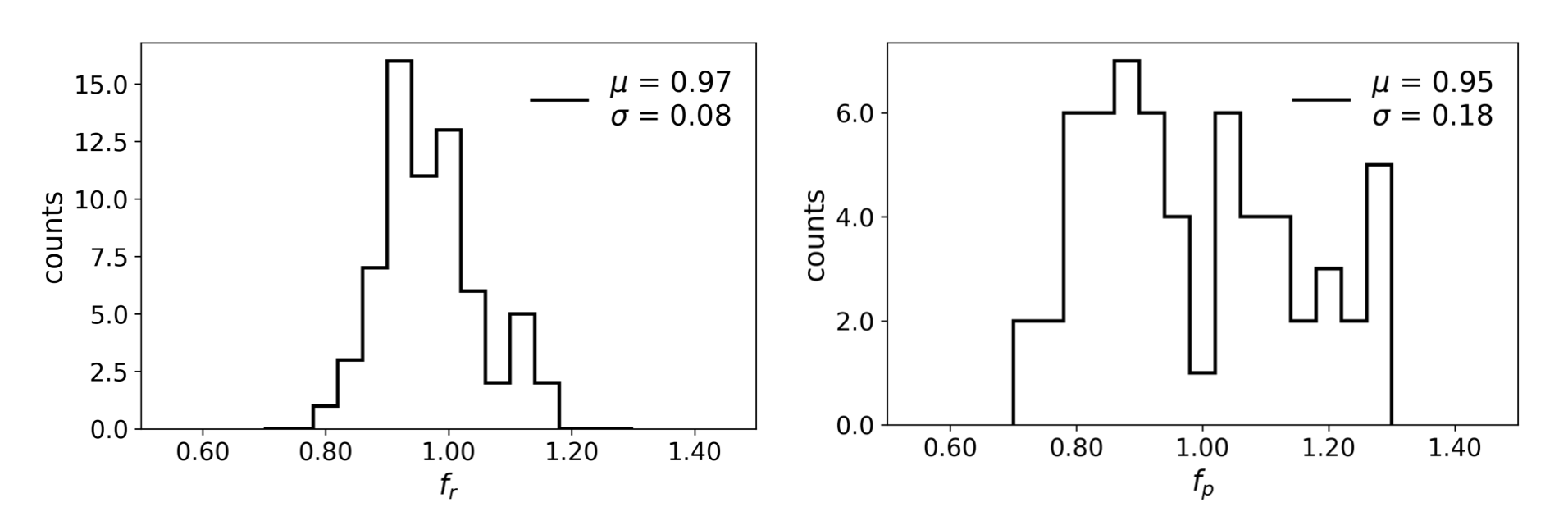}
\caption{Reconstructed scale factors resulting from a Monte Carlo vs. Monte Carlo study for one event.  Radio scale factors are shown in the left panel, and particle scale factors in the right panel.  The standard deviation of each distribution is taken as the fit uncertainty for this event.}
\label{fig:single_res}
\end{figure}

The reconstruction uncertainty of each event is derived using a Monte Carlo vs. Monte Carlo method.  The simulation set created for each event contains at least 40 showers.  One shower is used as mock ``data'' and the time-integrated power, $\varepsilon$, at the position of each LOFAR antenna is found using the two dimensional $\varepsilon_{\mathrm{sim}}$ map.  Noise is added to the ``data'' that reflects the noise level in the actual event data.  This ``event'' is then reconstructed using the remaining simulations using the same procedure discussed in Section~\ref{sec:energy_radio}.  The fit produces the scale factor $f_r$, used to convert the simulated energy $E_{\mathrm{sim}}$ to event energy as  $E_{\mathrm{radio}} = f_r \times E_{\mathrm{sim}}$.  Each shower in the set is simulated at the same energy, so if the fit were perfect the scale factor would always be $f_r=1$.  This procedure is repeated using every simulation in the set as the ``event'', yielding a set of scale factors, $f_r$.  The same procedure is also applied to the particle data, yielding a set of reconstructed scale factors $f_p$.  The resulting distributions of scale factors for one event are shown in the histograms in Fig.~\ref{fig:single_res}, where the radio scale factors are in the left panel and particle scale factors are in the right panel.  One standard deviation of the distribution is taken to be the fit uncertainty for a particular event.

\begin{figure}[ht]
\centering
\includegraphics[scale=0.45]{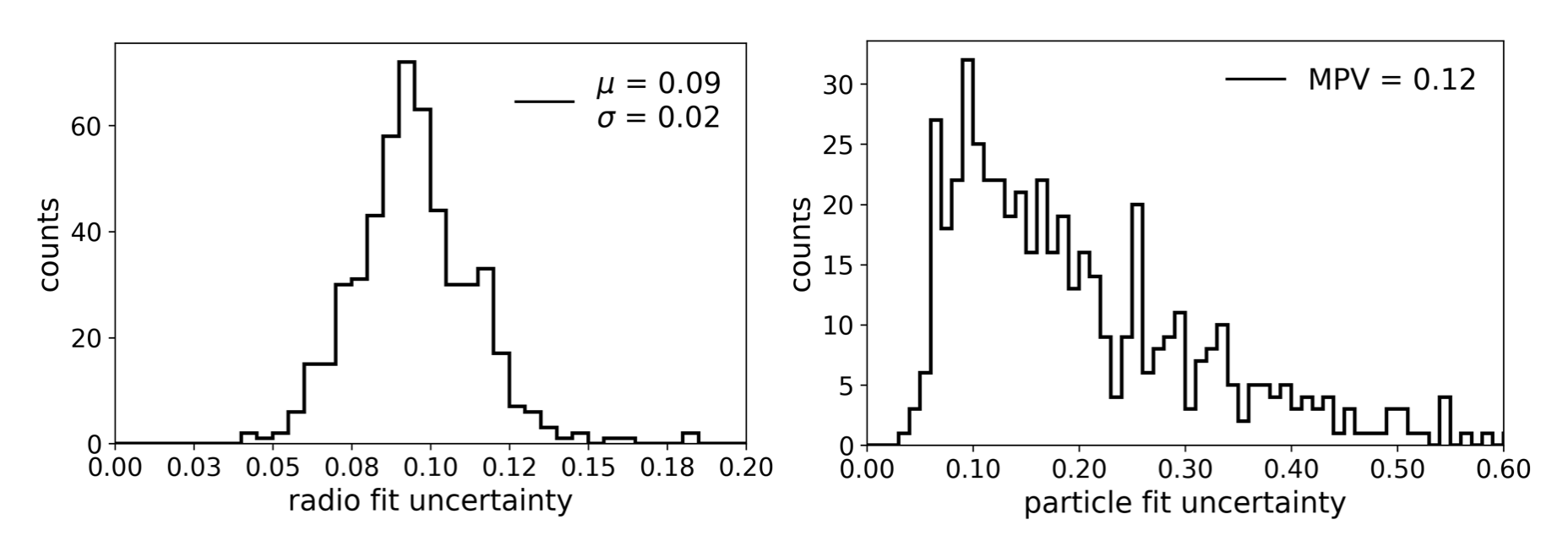}
\caption{Distribution of fit uncertainties for radio-based energy reconstruction (left), and particle-based energy reconstruction (right).}
\label{fig:all_res}
\end{figure}

Fig.~\ref{fig:all_res} shows the distribution of the standard deviation of scale factors for each event, or equivalently, the reconstruction uncertainty, for all events.  Again, the radio uncertainty is shown in the left panel, and particle uncertainty in the right.  We see that typical values for reconstruction uncertainties on radio-based energy are close to 9\% with little spread.  The most probable value (MPV) of the particle-based uncertainties is 12\%, but they have a much larger spread and extend to 50\%.

\vspace{4mm}
\noindent\textbf{Radio and Particle: composition uncertainties}
\vspace{4mm}

Although the radio reconstruction technique discussed in Section~\ref{sec:energy_radio} reconstructs $X_{\mathrm{max}}$ to within 17 g/cm$^2$~\cite{buitink2014}, there is still uncertainty associated with the primary composition.  For each event, a best fit simulation is chosen out of a set of more than 40 simulations of both proton and iron primaries.  The energy of the event is then found using equations \ref{eq:chi2} and \ref{eq:radio_energy}.  The fact that the best fit simulation is associated with either a proton or an iron primary affects the reconstructed energy.  In order to quantify this uncertainty, we ran simulations for a subset of events collecting a set of both 15 proton and 15 iron simulations with $X_{\mathrm{max}}$ values within 5 g/cm$^2$ of the reconstructed value of the event.   We found that iron-initiated showers consistently reconstruct with an radio-based energy 10\% higher than proton-initiated showers.  This effect is independent of $X_{\mathrm{max}}$ or zenith angle.  For each event, we include an asymmetric 10\% uncertainty.  For example, if the best fit simulation corresponds to a proton primary, the error bar is included in the positive direction.  This provides a conservative estimate for the composition uncertainty on the radio-based energy reconstruction.

Likewise, there is a composition uncertainty on the particle-based energy reconstruction. Given the same $X_{\mathrm{max}}$, proton-initiated showers reconstruct with a higher energy than iron-initiated showers.  This effect is amplified with increasing zenith angle.  For vertical showers, proton-initiated events reconstruct a few percent higher than iron.  At 50$^{\circ}$ zenith, this difference increases to 30\%.  We have parameterized the difference between proton and iron energy reconstruction as a function of zenith angle, and added asymmetric error bars accordingly.

\vspace{4mm}
\noindent\textbf{Radio: angular dependence of the antenna model}
\vspace{4mm}

\begin{figure}[ht]
\centering
\includegraphics[scale=0.6]{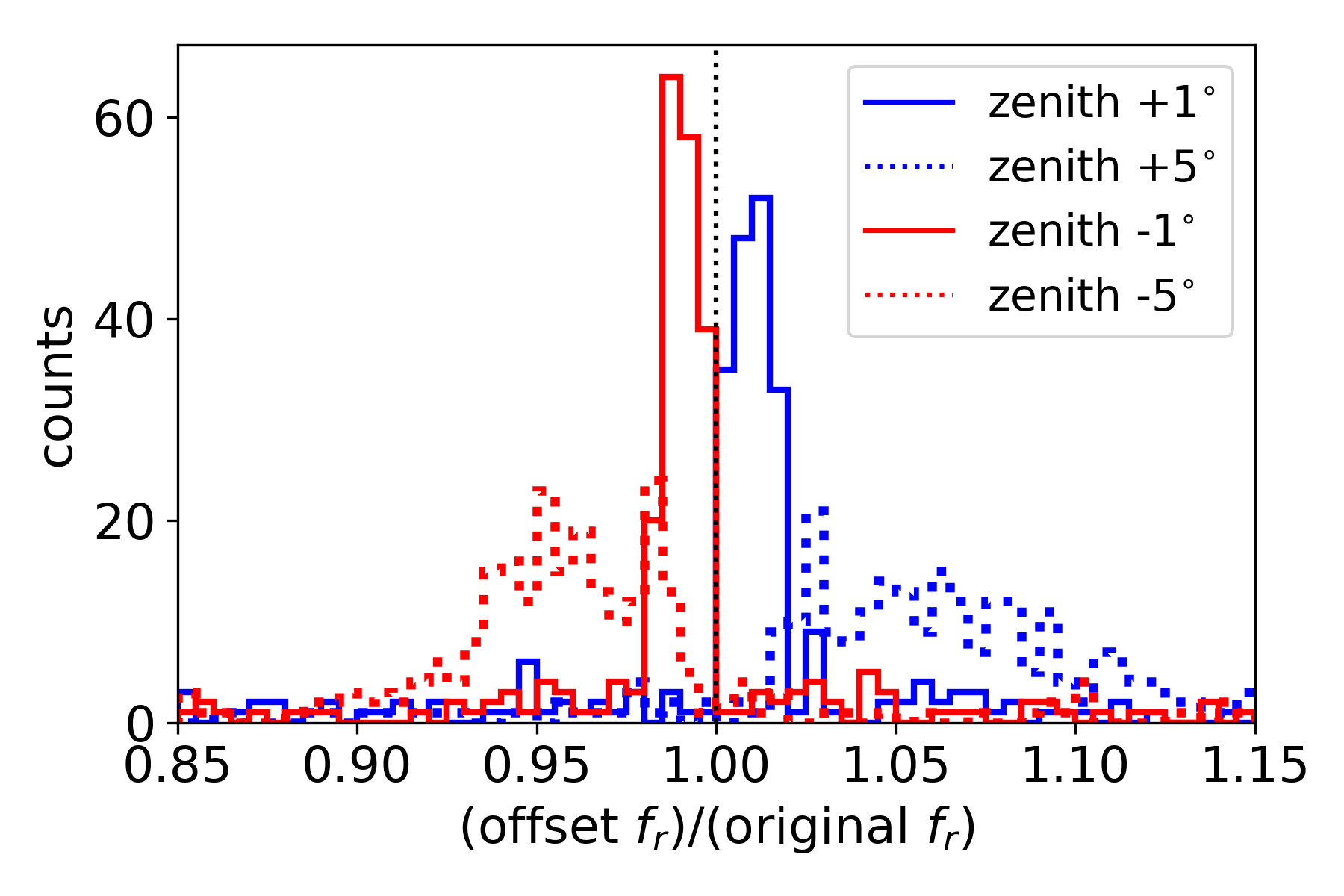}
\caption{The ratio between energy reconstructed with antenna model offsets and the original model.  Offsets of $\pm~1^{\circ}$ and $\pm~5^{\circ}$ were applied.}
\label{fig:reco_antenna_offset}
\end{figure}

We account for the uncertainty in the angular dependence of the antenna model.  The overall calibration of the antenna model is handled separately~\cite{Mulrey:2019vtz}.  Measurements have been made using a reference source attached to an octocopter that showed that the received power as a function of direction, based on the antenna model, is in general agreement with measurements~\cite{Nelles:2015gca}.  In order to estimate the remaining uncertainty, we offset the antenna model by $\pm~1^{\circ}$ and $\pm~5^{\circ}$ in the zenith direction.  Events are then reprocessed with the offset antenna model to determine the effect on reconstructed energy.  The ratio between the energy reconstructed with the offset and without the offset for each event is shown in Fig.~\ref{fig:reco_antenna_offset}.  We take an uncertainty of $\pm$~5\%, which is reflective of the distributions of the energy reconstruction ratio with the model offset $5^{\circ}$.  This is a conservative estimate, because in practice, each antenna may have a slightly different offset in different directions, in which case the effect of the antenna model will have less impact on the final energy.

\vspace{4mm}
\noindent\textbf{Radio: temperature dependence}
\vspace{4mm}

There was no apparent seasonal or temperature dependence found in the antenna gain.  This was determined by looking at the average power contained in the background of event traces over the course of the year.  The dominant contribution to the background radio signal is Galactic noise.  While the same sky is visible over 24 hours in local sidereal time, the portion that is visible during nighttime and daytime hours changes from winter to summer;  more Galactic emission is visible at night in the winter.  There are more LBA observations in night hours, and so the average power increases in the winter months.  The effect is that there is 10\% more noise power in the traces, but this effect is simply an offset and not multiplicative, and is handled in the data processing pipeline.  We also do not see any degradation in the average antenna signal over time.

\vspace{4mm}
\noindent\textbf{Particle: scintillator response variation}
\vspace{4mm}

The response of the scintillators varies over time.  We characterize the variation of each scintillator by looking at the daily average energy deposit.  The variation is then determined by taking the standard deviation of average energy deposits.  There is no apparent systematic variation with temperature, degradation over time, or correlation between detectors.  In order to propagate this uncertainty into the energy reconstruction, we repeat the analysis while adjusting the scintillator calibration, or equivalently, the conversion factor from deposited energy in ADC units into MeV.  Since the variation in each scintillator is uncorrelated, the adjustment to each scintillator's calibration is made by selecting a new value from a Gaussian distribution with the original conversion factor as the mean, and variation as the standard deviation.  This process was repeated three times, and the maximum effect on the resulting energy scale was 2.5\%.

\subsection{Systematic Uncertainties}

\vspace{4mm}
\noindent\textbf{Radio: antenna calibration}
\vspace{4mm}

The antenna and signal chain calibration is the dominant systematic uncertainty in the radio-based energy reconstruction at 13\%.  The calibration is based on modeling the propagation of Galactic noise through the LOFAR signal chain.  Details are given in~\cite{Mulrey:2019vtz}.  The dominant contribution to the uncertainty on the calibration is the uncertainty on the underlying models predicting the brightness temperature of the galaxy, which contribute 11\%.

\vspace{4mm}
\noindent\textbf{Radio: choice of simulation code}
\vspace{4mm}

The simulated radio emission for this analysis was generated using CoREAS.  CoREAS determines radio emission from a given particle track in the cascade by applying the endpoint formalism~\cite{Huege:2013vt,James:2010vm}.  There are other methods, including the approach taken in ZHAireS, which implements the ZHS method~\cite{Zas:1991jv,AlvarezMuniz:2010ty}.  The agreement between the two simulation codes was studied in~\cite{Gottowik:2017wio}, and the effect of choosing one code over the other on the energy scale was determined to be less than 2.6\%.

\vspace{4mm}
\noindent\textbf{Radio and particle: hadronic interaction model}
\vspace{4mm}

In order to determine the effect of the choice of hadronic interaction model used in the CORSIKA simulation, ten events, with a minimum of 40 simulations each, were re-analyzed using simulations produced with the Sibyll 2.3c~\cite{Fletcher:1994bd} interaction model instead of \mbox{QGSJETII-04}.  This changed the resulting radio-based energy by 3\%  and particle-based energy by 7\%.

\vspace{4mm}
\noindent\textbf{Particle: scintillator calibration}
\vspace{4mm}

\begin{figure}[ht]
\centering
\includegraphics[scale=0.6]{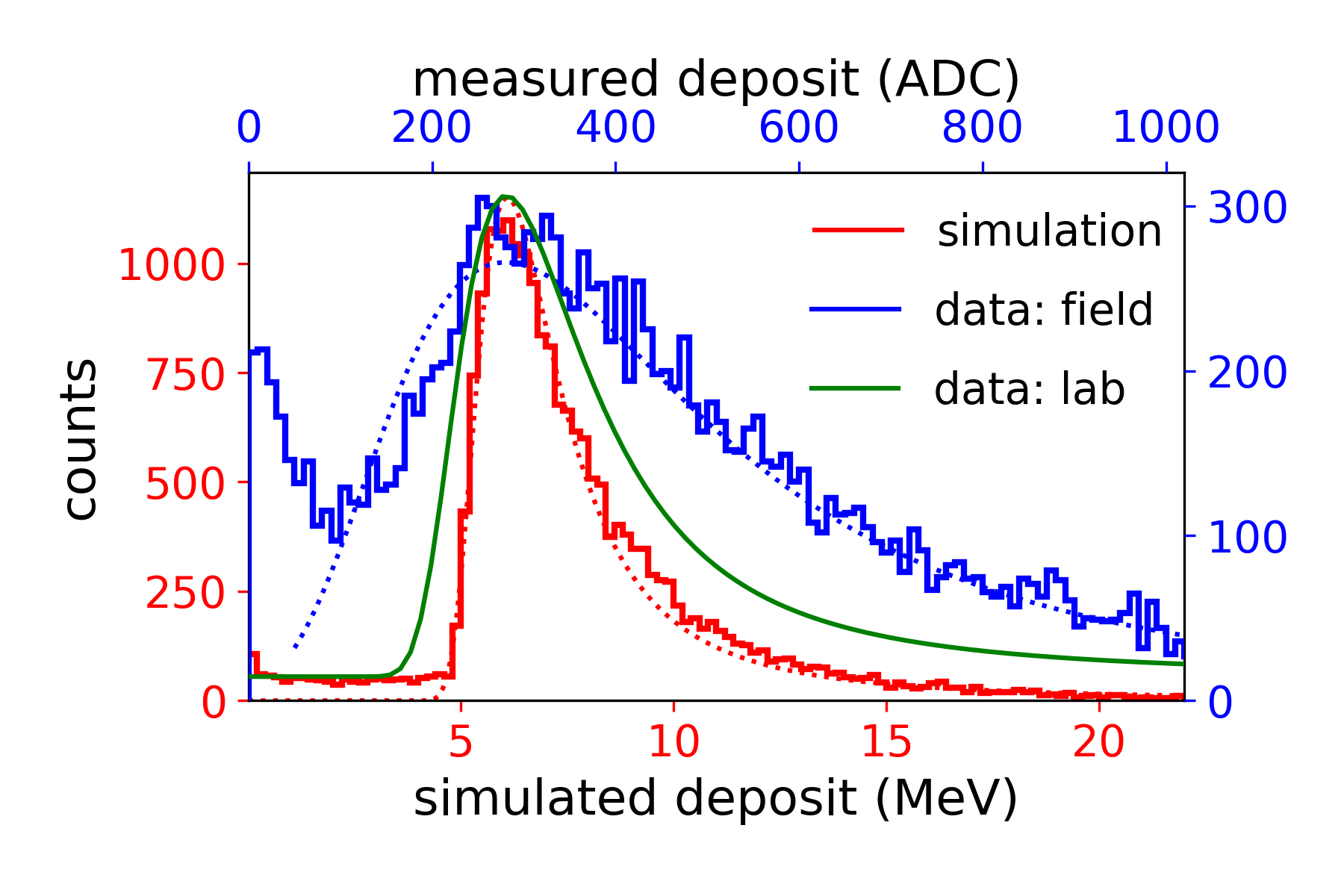}
\caption{Distributions of energy deposits from single muons.  The distribution in red shows the GEANT4 simulated deposits. The green and blue lines represent muon energy deposits measured in the lab and field, respectively.}
\label{fig:geant_muons}
\end{figure}

In order to calibrate the scintillators, or in other words, to find the conversion between charge deposit in ADC counts and MeV, it is necessary to collect single muon events.  This is done in the field by setting the trigger threshold very low, and collecting a combination of singly charged particle events (presumably dominated by muons) and noise triggers from a single scintillator.  Then, the most probable muon deposit from LORA data can be compared to the most probable energy deposit based on GEANT4 simulated muon events using a realistic detector description and arrival directions following a $\mathrm{cos}(\theta)^2$ zenith angle distribution~\cite{geant4}.  The resulting muon energy deposit distributions are shown in Fig.~\ref{fig:geant_muons}.  The distribution shown in red comes from simulation, and the blue distribution shows events measured in the field.  For the field events, the noise peak is also visible.  We also measured the energy deposit of single muon events in a laboratory setting, using a muon tower to ensure the triggered events were exclusively single muons.  The resulting distribution is shown in green.  The field measurements have a broader distribution due to the environment being less controlled, but the peak of the distribution still represents the most probably muon energy deposit.

\begin{figure}[ht]
\centering
\includegraphics[scale=0.45]{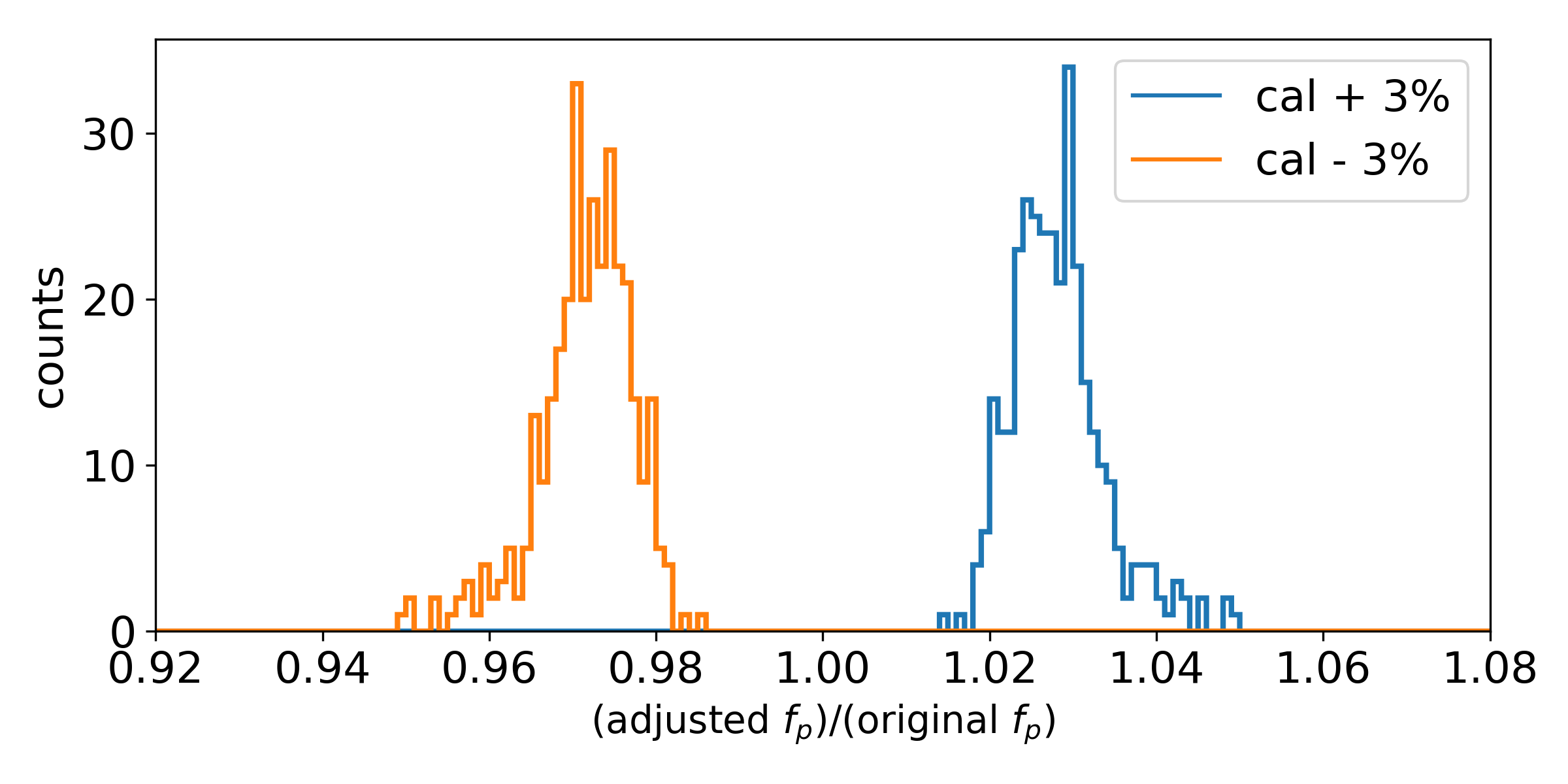}
\caption{Ratio of energy reconstructed with the standard calibration plus 3\% (in blue), and minus 3\% (in orange) to the original energy reconstruction.}
\label{fig:cal_sys}
\end{figure}

There is a systematic uncertainty associated with the calibration of the LORA scintillators.  By repeating the field calibration process a number of times, we find the standard deviation of the calibration values for each scintillator.  Then, this uncertainty is propagated through the analysis, and new energy reconstructions are obtained.  The average standard deviation of the scintillator calibration value is $\pm~3\%$, which propagates into $\pm~3\%$ uncertainty in event energy.  This is shown in Fig.~\ref{fig:cal_sys}.  The histograms represent the ratio of energy reconstructed with the standard calibration plus 3\% (in blue), and minus 3\% (in orange) to the original energy reconstruction.  

The spatial response of the scintillators was measured at Karlsruhe Institute of Technology (KIT) and using a muon tracking detector which was originally used for muon tracking in the KASCADE experiment~\cite{KASCADE_NIM}.  The response is very uniform over the surface of the scintillator, and is now included in LORA simulations~\cite{mulrey2019icrcLORA}.

\acknowledgments

We thank the referee for thoughtful feedback that lead to an improved analysis.  The LOFAR cosmic-ray key science project acknowledges funding from an Advanced Grant of the European Research Council (FP/2007-2013) / ERC Grant Agreement n. 227610. The project has also received funding from the European Research Council (ERC) under the European Union’s Horizon 2020 research and innovation programme (grant agreement No 640130). We furthermore acknowledge financial support from FOM, (FOM-project 12PR304).  ST acknowledges funding from the Khalifa University Startup grant (project code 8474000237). BMH is supported by NWO (VI.VENI.192.071). KM is supported by FWO (FWOTM944). AN acknowledges the DFG grant NE 2031/2-1. LOFAR, the Low Frequency Array designed and constructed by ASTRON, has facilities in several countries, that are owned by various parties (each with their own funding sources), and that are collectively operated by the International LOFAR Telescope foundation under a joint scientific policy.


\newpage
%\printbibliography[heading=subbibliography,notkeyword=this]
\bibliography{main}
%\bibliographystyle{plain}

%\begin{thebibliography}{99}

%\bibitem{a}
%Author, \emph{Title}, \emph{J. Abbrev.} {\bf vol} (year) pg.

%\bibitem{b}
%Author, \emph{Title},
%%arxiv:1234.5678.

%\bibitem{c}
%Author, \emph{Title},
%Publisher (year).

% Please avoid comments such as "For a review'', "For some examples",
% "and references therein" or move them in the text. In general,
% please leave only references in the bibliography and move all
% accessory text in footnotes.

% Also, please have only one work for each \bibitem.

%\end{thebibliography}
\end{document}